\documentclass[aps,preprint,12pt,noshowpacs,showkeys,nofootinbib,eqsecnum,hyperref]{revtex4}
\usepackage{amssymb}
\usepackage{amsfonts}
\usepackage{amsmath}
\usepackage{graphicx}

\input{tcilatex}

\begin{document}

\title{Is String Interaction the Origin of Quantum Mechanics? }
\author{Itzhak Bars and Dmitry Rychkov}
\author{}
\affiliation{Department of Physics and Astronomy, University of Southern California, Los
Angeles, CA 90089-0484, USA}

\begin{abstract}
String theory developed by demanding consistency with quantum mechanics. In
this paper we wish to reverse the reasoning. We pretend open string field
theory is a fully consistent definition of the theory - it is at least a
self consistent sector. Then we find in its structure that the rules of
quantum mechanics emerge from the non-commutative nature of the basic string
joining/splitting interactions. Thus, rather than assuming the quantum
commutation rules among the usual canonical variables we derive them from
the physical process of string interactions. Morally we could apply such an
argument to M-theory to cover quantum mechanics for all physics. If string
or M-theory really underlies all physics, it seems that the door has been
opened to an explanation of the origins of quantum mechanics from physical
processes.
\end{abstract}

\maketitle

\section{Introduction}

Quantum mechanics (QM) works amazingly well in all known parts of
microscopic physics. One can deduce classical physics as the limit of QM for
large quantum numbers (or equivalently the small $\hbar $ limit). Hence the
general belief is that QM is the only rule for all types of mechanics.
Despite the tremendous success of QM, the fundamental commutation rules from
which all QM is derived, namely $\left[ x,p\right] =i\hbar $ for every
degree of freedom, needs to be put in mysteriously \textquotedblleft by
hand\textquotedblright\ without any underlying reasoning. It is well
established that, if the quantization rule is accepted, then all of the
amazing and correct consequences of quantum mechanics follow. The success of
QM is of course a justification to accept the mysterious rule as correct,
but it leaves us begging for an underlying explanation.

In this paper we will present arguments that there may be a physical
explanation for where the QM rules come from. We will show that there is a
clear link between the commutation rules of QM operators and the
non-commutative string joining/splitting interactions \cite{witten} that
were expressed in the language of the Moyal star formulation of string field
theory (MSFT) \cite{B} in a recently improved and more intuitive version 
\cite{MSFT-BR1}. Except for the \textit{mathematical similarity}, the Moyal $%
\star $ in MSFT has nothing to do with the Moyal product \cite{Moyal} that
reproduces\footnote{%
For the explanation of how the well known Moyal product \cite{Moyal} for
classical phase space functions reproduces all the details of quantum
mechanics, read section (III) in \cite{MSFT-BR1} which summarizes the
essentials of this correspondence. \label{footMoyal}} QM, because the basic
non-commuting quantities in the string $\star $ in MSFT are very different
than the canonical conjugates indicated by quantum mechanics. Nevertheless,
we found how to link the basic QM commutators to the string $\star $ and
derive the \textit{QM rules only from the rules of string joining/splitting}%
. This link suggests that there is a deeper physical phenomenon, namely
string interactions, underlying the usual quantum rules of QM, thus
providing a possible explanation for where they come from.

The essential arguments for the thesis of this paper can be adequately
presented in a simplified model that captures the necessary ingredients of
MSFT. The simplified model, which we call mini-MSFT, consists basically of
the phase space system of two particles, rather than the full phase space of
an infinite number of particles that make up all the points on a string. The
two particles may be thought of as the end points of an open string, but it
is also possible not to think of the string concept at all to discuss the
main ideas. This is because only the properties of phase space, rather than
the property of the dynamics of the two particles enter in the main part of
the discussion. Hence to keep our discussion as simple as possible, we will
define the mini-MSFT system in section (\ref{mini}) and discuss how to
derive the QM properties from \textquotedblleft string\textquotedblright\
interactions. The mini-MSFT may be a useful model in its own right to
discuss some physically interesting systems, as in the examples we outline
at the end of section (\ref{mini}).

Even though we will not use the full machinery of MSFT in this paper, we
begin our discussion in section (\ref{msftDOFs}) with a brief description of
its setup so that the reader, even without knowing much about string theory,
can see the connection between the full string field theory and the
simplified 2-particle model in section (\ref{mini}), and be able to deduce
easily that the arguments for the thesis of this paper given in the context
of the simple model in section (\ref{mini}) apply equally well to the full
string theory in our preferred MSFT language for the full string. The full
string theory (and its M-theory extension) is needed to be able to apply the
argument to all physics, provided one is willing to make the assumption that
string- or M-theory actually underlies all physics.

\section{Degrees of freedom in MSFT \label{msftDOFs}}

The open string position degrees of freedom $X^{M}\left( \sigma \right) ,$
at a fixed value of the worldsheet parameter $\tau ,$ are parametrized by
the worldsheet parameter $\sigma $, with $0\leq \sigma \leq \pi $. Witten 
\cite{witten} suggested to imagine the string field $\psi \left( X\left(
\sigma \right) \right) $ as an infinite dimensional matrix $\psi _{ij}\left( 
\bar{x}\right) $ 
\begin{equation}
\psi \left( X\left( \sigma \right) \right) =\psi _{x_{L}\left( \sigma
\right) ,x_{R}\left( \sigma \right) }\left( \bar{x}\right) ,  \label{Mpsi}
\end{equation}%
whose left/right labels $i\sim x_{L}^{M}\left( \sigma \right) ,$ and $j\sim
x_{R}^{M}\left( \sigma \right) $ are the left/right halves of the string
relative to the midpoint at $\sigma =\pi /2,$ namely $x_{L}^{M}\left( \sigma
\right) =\left\{ X^{M}\left( \sigma \right) \text{ for }0\leq \sigma <\pi
/2\right\} $ and $x_{R}^{M}\left( \sigma \right) =\left\{ X^{M}\left( \sigma
\right) \text{ for }\pi /2<\sigma \leq \pi ,\right\} ,$ while $\bar{x}%
^{M}\equiv X^{M}\left( \pi /2\right) $ is the location of the midpoint. It
was then suggested in \cite{witten} that the products of fields $\psi
_{1}\left( X\right) \star \psi _{2}\left( X\right) =\psi _{12}\left(
X\right) $ in open string field theory is the non-commutative matrix product
of matrices of the form (\ref{Mpsi}), and that the action is similar to the
Chern-Simons theory 
\begin{equation}
S=\int \left( \frac{1}{2}\psi \star \left( \hat{Q}\psi \right) +\frac{g}{3}%
\psi \star \psi \star \psi \right) ,  \label{cubicW}
\end{equation}%
where $\hat{Q}$ is the BRST operator of a conformal field theory on the
worldsheet (CFT). This proposal worked and produced correctly the Veneziano
model type perturbative string scattering amplitudes \cite{giddings}.

The matrix product in $\psi \star \psi $ was implemented by going back to
the worldsheet conformal field theory to perform computations, which proved
to be prohibitively complicated and took away much of the simplicity and
elegance of the matrix-like setup of the $\star $ product and the action.
Seeking a way of avoiding the complicated CFT maps, while keeping the
elegant algebraic structure, the Moyal star product formulation of string
Field theory (MSFT) was suggested in \cite{B}, and computations were
performed in \cite{BM}-\cite{BP}, showing that this was a more efficient
approach to compute and correctly recover the perturbative Veneziano
amplitudes, including a higher degree of accuracy for the off-shell versions
of the amplitudes \cite{BP}. The MSFT formalism has been reformulated
recently in \cite{MSFT-BR1} in a new basis of degrees of freedom in which
all expressions, especially the $\star $ product and computations greatly
simplify. It is the new form of the star product displayed below that
suggests the connection between string joining and quantum mechanics.

In the new version of MSFT the string field is taken to be a functional $%
A\left( x_{+},p_{-}\right) $ of \textit{half of the phase space} of the
string, where $x_{+}^{M}\left( \sigma \right) $ is the symmetric part of $%
X^{M}\left( \sigma \right) $ under reflections relative to the midpoint, $%
x_{+}^{M}\left( \sigma \right) =\frac{1}{2}\left( X^{M}\left( \sigma \right)
+X^{M}\left( \pi -\sigma \right) \right) ,$ while $p_{-M}\left( \sigma
\right) =\frac{1}{2}\left( P_{M}\left( \sigma \right) -P_{M}\left( \pi
-\sigma \right) \right) $ is the antisymmetric part of the momentum density.
Note that $p_{-M}\left( \sigma \right) $ is the canonical conjugate to $%
x_{-}^{M}\left( \sigma \right) $ and commutes with $x_{+}^{M}\left( \sigma
\right) $ in the first quantization of the string. The
symmetric/antisymmetric $x_{\pm }\left( \sigma \right) $ are related to $%
\left( x_{L},\bar{x},x_{R}\right) $ in the Witten version by $x_{\pm }=\frac{%
1}{2}\left( x_{L}\pm x_{R}\right) ,$ and including the midpoint $\bar{x}$ as
part of $x_{+}\left( \sigma \right) .$ Thus the MSFT field $A\left(
x_{+},p_{-}\right) $ is related to the field $\psi \left( X\right) =\psi
\left( x_{L},\bar{x},x_{R}\right) =\psi \left( x_{+},x_{-}\right) $ by a
Fourier transform from $x_{-}~$to $p_{-}.$ With this choice of \textit{%
half-phase-space degrees freedom} to label the string field $A\left(
x_{+}\left( \sigma \right) ,p_{-}\left( \sigma \right) \right) $, the
matrix-like product for string joining in position space $\psi _{12}\left(
X\right) =\psi _{1}\left( X\right) \star \psi _{2}\left( X\right) $ is
mapped to the Moyal product in the \textit{half-phase-space}, $A_{12}\left(
x_{+},p_{-}\right) =A_{1}\left( x_{+},p_{-}\right) \star A_{2}\left(
x_{+},p_{-}\right) $ with 
\begin{equation}
\star ~=\exp \left[ \frac{i}{4}\int_{0}^{\pi }d\sigma ~\text{sign}\left( 
\frac{\pi }{2}-\sigma \right) \left( \overrightarrow{\partial }%
_{p_{-M}\left( \sigma \right) }\overleftarrow{\partial }_{x_{+}^{M}\left(
\sigma ,\varepsilon \right) }-\overleftarrow{\partial }_{p_{-M}\left( \sigma
\right) }\overrightarrow{\partial }_{x_{+}^{M}\left( \sigma ,\varepsilon
\right) }\right) \right] .  \label{msftstar}
\end{equation}%
A very important property of the new star product is that it is background
independent because phase space does not care which conformal field theory
on the worldsheet underlies the string action $S_{string}$ or which
background fields it contains. The sum over the $M$ indices in (\ref%
{msftstar}) does not involve a metric because $X^{M}$ is defined with an
upper index and then $P_{M},$ which is derived from the action according to
the canonical procedure, $P_{M}=\partial S_{string}/\left( \partial _{\tau
}X^{M}\right) ,$ automatically has a lower index.

An elegant aspect of MSFT that will be centrally relevant for our discussion
in this paper is that the quantum canonical operators for any point on the
string $\hat{X}\left( \sigma \right) ,\hat{P}\left( \sigma \right) $ are
represented on the string field $A\left( x_{+},p_{-}\right) $ \textit{only
by string joining/splitting operations}, namely by either left or right
star-multiplication, depending on whether the point is to the left or to the
right of the midpoint at $\sigma =\pi /2$ 
\begin{align}
\hat{X}^{M}\left( \sigma ,\varepsilon \right) A\left( x_{+},p_{-}\right) &
=\left\{ 
\begin{array}{l}
x_{+}^{M}\left( \sigma ,\varepsilon \right) \star A\left( x_{+},p_{-}\right)
,\text{ if }0\leq \sigma \leq \pi /2 \\ 
A\left( x,p\right) \star x_{+}\left( \sigma ,\varepsilon \right) \left(
-1\right) ^{MA},\text{ if }\pi /2\leq \sigma \leq \pi%
\end{array}%
\right. ,  \label{iQMX2} \\
\hat{P}_{M}\left( \sigma \right) \left( x_{+},p_{-}\right) & =\left\{ 
\begin{array}{l}
\left( e^{-\varepsilon \left\vert \partial _{\sigma }\right\vert
}p_{-M}\left( \sigma \right) \right) \star A\left( x_{+},p_{-}\right) ,\text{
if }0\leq \sigma \leq \pi /2 \\ 
A\left( x_{+},p_{-}\right) \star \left( e^{-\varepsilon \left\vert \partial
_{\sigma }\right\vert }p_{-M}\left( \sigma \right) \right) \left( -1\right)
^{MA},\text{ if }\pi /2\leq \sigma \leq \pi%
\end{array}%
\right. .  \label{iQMP2}
\end{align}%
Note that on the right hand side the string fields that are being joined are 
$A\left( x_{+},p_{-}\right) $ and $x_{+}$ or $A$ and $p_{-},$ where $%
x_{+},p_{-}$ are specialized cases of a more general string field $A\left(
x_{+},p_{-}\right) .$

This $\star $ product includes a small parameter $\varepsilon $ which is a
regulator to avoid notorious midpoint anomalies, and the label $M=\left( \mu
,b,c\right) $ includes spacetime $\left( \mu \right) $ and ghost $\left(
b,c\right) $ degrees of freedom, all of which are necessary and insure a
well defined theory. For the reader interested in the details we suggest 
\cite{MSFT-BR1}. None of these complications will be needed to discuss the
main points of this paper. We will switch to the mini-MSFT that imitates in
a simplified way only the star product for string splitting/joining using
only two particles. The remainder of this paper should be understandable to
the reader without having to know anything about strings or string field
theory.

\section{Toy Model with Two Particles (mini-MSFT) \label{mini}}

We begin with the phase space of two particles named $L$ (left) and $R$
(right). It may be helpful to imagine that these correspond to the two end
points of a string, however this picture is not necessary and the setup
below may apply to more general physical circumstances. The particles are
located at arbitrary positions ($\vec{x}_{L},\vec{x}_{R}),$ and have
canonical conjugate momenta $(\vec{p}_{L},\vec{p}_{R}).$ Their center of
mass and relative coordinates are, $R^{i}=\frac{1}{2}\left(
x_{L}^{i}+x_{R}^{i}\right) ,$ and $r^{i}=\left( x_{L}^{i}-x_{R}^{i}\right) ,$
while the momenta canonically conjugate to $(R^{i},r^{i}$) are the total
momentum $P_{i}=\left( p_{iL}+p_{iR}\right) ,$ and the relative momentum $%
p_{i}=\frac{1}{2}\left( p_{iL}-p_{iR}\right) .$ The dynamics is controlled
by some Hamiltonian $H(P,p,R,r)$ whose details are unimportant for now%
\footnote{%
The Hamiltonian $H$ in the toy model is the analog of the Virasoro operator $%
L_{0}$ for a string in any background, that plays a role of the kinetic
energy operator in the quadratic term in string field theory in the Siegel
gauge. More generally, the kinetic operator in string field theory is the
BRST operator as in (\ref{cubicW}).}. Independent of the Hamiltonian, the
phase space $(P,p,R,r)$ has the standard canonical properties, namely we may
define classical Poisson brackets or quantum commutators based on the
canonical pairs $(R^{i},P_{j})$ and $(r^{i},p_{j})$. In particular, the
classical Poisson bracket between any two phase space functions, $%
U(P,p,R,r),V(P,p,R,r),$ is 
\begin{equation}
\left\{ U,V\right\} =\frac{\partial U}{\partial R^{i}}\frac{\partial V}{%
\partial P_{i}}-\frac{\partial U}{\partial P_{i}}\frac{\partial V}{\partial
R^{i}}+\frac{\partial U}{\partial r^{i}}\frac{\partial V}{\partial p_{i}}-%
\frac{\partial U}{\partial p_{i}}\frac{\partial V}{\partial r^{i}}.
\label{poisson}
\end{equation}

To proceed with usual quantization in quantum mechanics (QM) we may define
the eigenspace basis for a complete set of commuting operators, such as
position space, $\langle \vec{x}_{L},\vec{x}_{R}|$ or $\langle \vec{R},\vec{r%
}|,$ and express the probability amplitude for an arbitrary quantum state $%
|\psi \rangle $ in any such basis as the dot product in the Hilbert space,
e.g. $\psi \left( x_{L},x_{R}\right) =\langle x_{L},x_{R}|\psi \rangle $ or $%
\psi \left( R,r\right) =\langle R,r|\psi \rangle .$ We will be interested in
the Fourier transform of the latter 
\begin{equation}
A\left( R,p\right) \overset{Fourrier~\left( p,r\right) }{\leftrightarrow }%
\psi \left( R,r\right) .  \label{fourrier}
\end{equation}%
where $\langle R,p|$ is the complete eigenbasis for the space of the
commuting operators ($\hat{R}^{i},\hat{p}_{i}).$ We will think of the
probability amplitude, $A\left( R,p\right) =\langle R,p|\psi \rangle ,$ as a
field in a field theory as a function of \textit{the classical
half-phase-space} ($\vec{R},\vec{p}$). This setup is motivated by MSFT that
was briefly outlined in section (\ref{msftDOFs}). We will call the toy model
in this section \textquotedblleft mini-MSFT\textquotedblright . The
parallels between the full MSFT and mini-MSFT are 
\begin{equation*}
R^{i}\sim x_{+}^{M}\left( \sigma \right) ,\;r^{i}\sim x_{-}^{M}\left( \sigma
\right) ,\;P_{i}\sim p_{+M}\left( \sigma \right) ,\;p_{i}\sim p_{-M}\left(
\sigma \right)
\end{equation*}%
and we did not care to make parallels between $i$ and $M,$ which permits
many possibilities including bosons and fermions (see \cite{MSFT-BR1}), but
to keep the discussion simple it is sufficient to consider bosonic Euclidean
space for $i.$

To quantize this 2-particle system in a new way we will take the approach
inspired by MSFT. We will not \`{a} priori assume the quantum commutation
rules of the operators $(\hat{P}_{i},\hat{p}_{i},\hat{R}^{i},\hat{r}^{i})$
that describe nature so well, but whose fundamental origin remains
mysterious. Rather, as the primary physical origin of QM we will begin from
a non-commutative product that has physical significance as interactions of
strings by joining/splitting. Only from the algebra of string
joining/splitting we will derive the quantum algebra of the operators $(\hat{%
P}_{i},\hat{p}_{i},\hat{R}^{i},\hat{r}^{i})$ without assuming it. String
joining/splitting was formulated for open strings in \cite{witten} as a
matrix-like product for the field as in (\ref{Mpsi}). For the present toy
model with only two particles we define a similar matrix-like product of
fields in position space in the form 
\begin{equation}
\psi _{12}\left( x_{L},x_{R}\right) =\int_{-\infty }^{\infty }d^{n}z~\psi
_{1}\left( x_{L},z\right) \psi _{2}\left( z,x_{R}\right) ,
\label{wittenstar1}
\end{equation}%
where each field $\psi \left( x_{L},x_{R}\right) $ is regarded as an
infinite dimensional matrix whose rows and columns are labelled by the
continuous indices $\left( x_{L},x_{R}\right) $ that correspond to the
locations of the two particles. The matrix-like rule (\ref{wittenstar1}) is
interpreted as a prescription for computing the probability amplitude $\psi
_{12}\left( x_{L},x_{R}\right) $ when two 2-particle clouds, described by $%
\psi _{1}\left( x_{L},x_{R}\right) $ and $\psi _{2}\left( x_{L},x_{R}\right)
,$ join together into a single cloud $\psi _{12}\left( x_{L},x_{R}\right) ,$
by annihilating a pair of particles, one from each cloud, when they meet
locally at all possible points $\vec{z}$ in the full volume. This is similar
to the picture for joining/splitting worldsheets, but in the present case
there are dynamical degrees of freedom only at the ends of the string. It
was shown in \cite{B}\cite{MSFT-BR1} that this string-like joining/splitting
can equivalently be formulated as a Moyal-type product, $A_{12}=A_{1}\star
A_{2},$ in the half-phase-space $\left( R^{i},p_{i}\right) $ related to
position space $\left( x_{L},x_{R}\right) $ by the Fourier transform
indicated in (\ref{fourrier}).

We now give the details of the $\star $ product in the half-phase space for
this simplified mini-MSFT. It is physically different but mathematically
analogous to the usual Moyal product:%
\begin{equation}
A_{12}\left( R,p\right) =\left( A_{1}\star A_{2}\right) \left( R,p\right)
=A_{1}\left( R,p\right) \exp \left( \frac{i\theta }{2}\left( \overleftarrow{%
\partial }_{R^{i}}\overrightarrow{\partial }_{p_{i}}-\overrightarrow{%
\partial }_{R^{i}}\overleftarrow{\partial }_{p_{i}}\right) \right)
A_{2}\left( R,p\right) .  \label{moyal}
\end{equation}%
It is the parallel of the string star product in (\ref{msftstar}). The
parameter $\theta $ must have the dimensions of the Planck constant $\hbar ,$
so it must be a multiple of $\hbar $ up to a dimensionless constant. In
fact, we will show that it is identically the Planck constant. The arrows in
(\ref{moyal}) instructs the reader to apply the derivatives on the functions
to the left ($A_{1}$) or right ($A_{2}$). For example, expanding in powers
of $\theta $ this $\star $ product gives%
\begin{equation}
A_{12}=A_{1}A_{2}+\frac{i\theta }{2}\left( \frac{\partial A_{1}}{\partial
R^{i}}\frac{\partial A_{2}}{\partial p_{i}}-\frac{\partial A_{1}}{\partial
p_{i}}\frac{\partial A_{2}}{\partial R^{i}}\right) +\cdots
\label{firstOrder}
\end{equation}%
The first order term in $\theta $ looks like a Poisson bracket, but this is
clearly different than the canonical Poisson bracket of classical mechanics
in Eq.(\ref{poisson}) since it does not involve the traditional canonical
conjugates exhibited in (\ref{poisson}). Instead, the center of mass
position $R^{i}$ and the relative momentum $p_{i},$ which belong to
different traditional canonical pairs, are set to play a new role analogous
to canonical conjugates in the half-phase-space ($R^{i},p_{i}$).

Using (\ref{moyal}) we compute $A_{1}\star A_{2}$ for the special cases when 
$A_{1}$ or $A_{2}$ is just $R^{i}$ or $p_{i},$ thus obtaining the left or
right multiplication of the general $A$ by the elementary degrees of freedom
in the half-phase-space 
\begin{equation}
\begin{array}{cc}
R^{i}\star A=\left( R^{i}+\frac{i\theta }{2}\frac{\overrightarrow{\partial }%
}{\partial p_{i}}\right) A\left( R,p\right) ,\;\; & A\star R^{i}=A\left(
R,p\right) \left( R^{i}-\frac{i\theta }{2}\frac{\overleftarrow{\partial }}{%
\partial p_{i}}\right) \\ 
p_{i}\star A=\left( p_{i}-\frac{i\theta }{2}\frac{\overrightarrow{\partial }%
}{\partial R^{i}}\right) A\left( R,p\right) ,\;\; & A\star p_{i}=A\left(
R,p\right) \left( R^{i}+\frac{i\theta }{2}\frac{\overleftarrow{\partial }}{%
\partial R^{i}}\right)%
\end{array}
\label{starLR}
\end{equation}%
There are no higher powers of $\theta $ because the higher derivatives in
the expansion of the exponential in (\ref{moyal}) vanish for this
computation. Other useful equivalent ways of writing the general $\star $
product are 
\begin{equation}
A_{1}\star A_{2}=\left\{ 
\begin{array}{c}
\left. A_{1}\left( \left( R^{\prime }+\frac{i\theta }{2}\overrightarrow{%
\partial }_{p}\right) ,\left( p^{\prime }-\frac{i\theta }{2}\overrightarrow{%
\partial }_{R}\right) \right) A_{2}\left( R,p\right) \right\vert _{R^{\prime
}=R,p^{\prime }=p} \\ 
\text{or} \\ 
\left. A_{1}\left( R^{\prime },p^{\prime }\right) A_{2}\left( \left( R-\frac{%
i\theta }{2}\overleftarrow{\partial }_{p^{\prime }}\right) ,\left( p+\frac{%
i\theta }{2}\overleftarrow{\partial }_{R^{\prime }}\right) \right)
\right\vert _{R^{\prime }=R,p^{\prime }=p}%
\end{array}%
\right.  \label{A12easy}
\end{equation}

Just like the well known Moyal star product \cite{Moyal}, which is related
to the Poisson bracket (\ref{poisson}) in the full phase space $(P,p,R,r)$,
reproduces all aspects of ordinary quantum mechanics (see footnote (\ref%
{footMoyal})), the string-joining Moyal star product in (\ref{moyal}) will
evidently produce a quantum-like system in the half-phase space $\left(
R,p\right) ,$ which we call \textit{induced quantum mechanics} (iQM)$.$ This
induced iQM has the following properties

\begin{itemize}
\item The product is associative $A_{1}\star\left( A_{2}\star A_{3}\right)
=\left( A_{1}\star A_{2}\right) \star A_{3}=A_{1}\star A_{2}\star A_{3},$
just as should be expected for the associative product of operators in the
induced iQM, where any product $\hat{A}_{1}\hat{A}_{2}\hat{A}_{3}\cdots$
does not require parentheses to be computed unambiguously.

\item By using (\ref{starLR}) we compute the products of the
half-phase-space elementary degrees of freedom $\left( R,p\right) $ 
\begin{equation}
\begin{array}{c}
R^{i}\star R^{j}=R^{i}R^{j},\;\;p_{i}\star p_{j}=p_{i}p_{j}, \\ 
R^{i}\star p_{j}=R^{i}p_{j}+\frac{i\theta }{2}\delta _{j}^{i},\;\;p_{j}\star
R^{i}=p_{j}R^{i}-\frac{i\theta }{2}\delta _{j}^{i}.%
\end{array}
\label{elementaryProducts}
\end{equation}%
This leads to the star commutator 
\begin{equation}
\left[ R^{i},p_{j}\right] _{\star }\equiv R^{i}\star p_{j}-p_{j}\star
R^{i}=i\theta \delta _{j}^{i}.  \label{iQMcommuteRules}
\end{equation}%
Hence $\left( R^{i},p_{j}\right) $ behave just like quantum mechanical
degrees of freedom. But \textit{this is not quantum mechanics} since in
ordinary QM the corresponding operators commute $\left[ \hat{R}^{i},\hat{p}%
_{j}\right] =0.$ Instead, this is the basic commutation property in the
induced iQM that comes from the non-commutative interactions in string
theory.
\end{itemize}

We now show that this induced iQM is a seed for constructing the usual QM in
the full operator space $(\hat{x}_{L}^{i},\hat{p}_{Li},\hat{x}_{R}^{i},\hat{p%
}_{Ri}).$ A map between operators in QM and their representative in iQM is
an elegant and intuitive property of MSFT as given in Eqs.(\ref{iQMX2},\ref%
{iQMP2}). Translated to mini-MSFT, his map is given only in terms of the $%
\star $ between \textit{two fields in the half-phase-space}, as follows 
\begin{equation}
\begin{array}{l}
\left. 
\begin{array}{l}
\hat{x}_{L}^{i}A=R^{i}\star A \\ 
\hat{p}_{iL}A=p_{i}\star A%
\end{array}%
\right\} \text{~ for particle }L\text{ the}\star \text{ product from left }
\\ 
\left. 
\begin{array}{l}
\hat{x}_{R}^{i}A=A\star R^{i} \\ 
\hat{p}_{iR}A=A\star \left( -p_{i}\right)%
\end{array}%
\right\} \text{~ for particle }R\text{ the}\star \text{ product from right}%
\end{array}
\label{QMrepresentation}
\end{equation}%
The reason for the $\left( -\right) $ sign in the last line is naturally
explained in the stringy version of the $\star $ in the full MSFT: it is
because for strings $x_{+}^{i}\left( \sigma \right) $ is symmetric with
respect to reflections from the midpoint, while $p_{-}\left( \sigma \right) $
is antisymmetric, leading to, $+p_{-}\left( \sigma \right) |_{\sigma \geq
\pi /2}\rightarrow -p_{-}\left( \sigma \right) |_{\sigma \leq \pi /2}.$
Using this map, let us now check the consistency between the commutation
rules in QM versus the iQM representatives above. We compute the commutators
by using only the $\star $ rules in (\ref{QMrepresentation}), associativity
of the $\star ,$ and the result for the $\star $ commutator in (\ref%
{iQMcommuteRules}). We find 
\begin{align}
\left[ \hat{x}_{L}^{i},\hat{p}_{Lj}\right] A& =\left[ R^{i},p_{j}\right]
_{\ast }\star A=i\theta \delta _{j}^{i}A,  \label{xpLR1} \\
\left[ \hat{x}_{R}^{i},\hat{p}_{Rj}\right] A& =A\star \left[ -p_{j},R^{i}%
\right] _{\ast }=i\theta \delta _{j}^{i}A,  \label{xpLR2} \\
\left[ \hat{x}_{L}^{i},\hat{p}_{Rj}\right] A& =-R^{i}\star A\star
p_{j}+R^{i}\star A\star p_{j}=0,  \label{xpLR3} \\
\left[ \hat{x}_{R}^{i},\hat{p}_{Lj}\right] A& =p_{j}\star A\star
R^{i}-p_{j}\star A\star R^{i}=0.  \label{xpLR4}
\end{align}%
For this iQM result to match the QM commutators of operators, $\left[ \hat{x}%
_{L}^{i},\hat{p}_{Lj}\right] =i\hbar \delta _{j}^{i}=\left[ \hat{x}_{R}^{i},%
\hat{p}_{Rj}\right] ,$ we must identify the parameter $\theta $ with the
Planck constant%
\begin{equation}
\theta =\hbar .
\end{equation}%
We have thus derived the basic rules of QM for each particle from the
string-interaction-induced iQM, by using only products of fields in the
half-phase-space, which signify string joining/splitting. Thus, the
non-commutativity inherent in the string interactions is directly connected
to the previously unexplained mysterious quantization rules of QM. So far
this is within a toy model, but since the same phenomenon is also true for
the full string theory (see \cite{MSFT-BR1}), assuming string theory is the
fundamental theory for all physics, then it becomes a statement for all
physics.

Continuing with mini-MSFT, next we investigate some operators constructed
from the basic ones. From the basic properties in (\ref{QMrepresentation})
we may extract the representation of each operator $(\hat{P}_{i},\hat{p}_{i},%
\hat{R}^{i},\hat{r}^{i}),$ in terms of only the $\star $ product of fields,
and then evaluate the star products in each line below by using (\ref{starLR}%
), after inserting $\theta =\hbar ,$ as follows 
\begin{equation}
\begin{array}{l}
\hat{R}^{i}A=\frac{1}{2}\left( \hat{x}_{L}^{i}+\hat{x}_{R}^{i}\right) A=%
\frac{1}{2}\left( R^{i}\star A+A\star R^{i}\right) =R^{i}A~, \\ 
\hat{r}^{i}A=\left( \hat{x}_{L}^{i}-\hat{x}_{R}^{i}\right) A=\left(
R^{i}\star A-A\star R^{i}\right) =i\hbar \partial _{p_{i}}A~, \\ 
\hat{P}_{i}A=\left( \hat{p}_{iL}+\hat{p}_{iR}\right) A=\left( p_{i}\star
A-A\star p_{i}\right) =-i\hbar \partial _{R^{i}}A~, \\ 
\hat{p}_{i}A=\frac{1}{2}\left( \hat{p}_{iL}-\hat{p}_{iR}\right) A=\frac{1}{2}%
\left( p_{i}\star A+A\star p_{i}\right) =p_{i}A~.%
\end{array}
\label{representation}
\end{equation}%
The end result in terms of differential operator representation is fully
consistent with the corresponding well known differential operator
representation of operators in QM. But the point here is that this result
follows from only the string joining/splitting interactions via the $\star $
product of fields given in (\ref{moyal}) and (\ref{QMrepresentation}).

Going further, from (\ref{QMrepresentation}) we derive the following
additional nice results which were significant in the formulation of MSFT 
\cite{MSFT-BR1}: if we have any quantum operator $\hat{O}_{L}\left( \hat{x}%
_{L},\hat{p}_{L}\right) $ (similarly $\hat{O}_{R}\left( \hat{x}_{R},\hat{p}%
_{R}\right) $) in usual QM, constructed from only the degrees of freedom of
particle $L$ (similarly $R$)$,$ then its representation in the iQM version
is given by the same function in which we replace ($\hat{x}%
_{L}^{i}\rightarrow R^{i}\star $ and $\hat{p}_{iL}\rightarrow p_{i}\star $)
and similarly ($\hat{x}_{R}^{i}\rightarrow \star R^{i}$ and $\hat{p}%
_{iR}\rightarrow \star \left( -p_{i}\right) $), where the $\star $ is to the
right (left) of $R$ or $p.$ Namely 
\begin{equation}
\begin{array}{c}
\hat{O}_{L}\left( \hat{x}_{L},\hat{p}_{L}\right) A=O_{L\star }\left(
R,p\right) \star A\;,\;\;\text{(}\star ~\text{from left).} \\ 
\hat{O}_{R}\left( \hat{x}_{R},\hat{p}_{R}\right) A=A\star O_{\star R}\left(
R,-p\right) \;,\;\;\text{(}\star ~\text{from right).}%
\end{array}
\label{analogLRmidpoint}
\end{equation}%
where $O_{L\star }$ means that all $R,p$ factors within it are star
multiplied with each other in the same order that operators appear in the QM
version, while in the case of $O_{\star R}$ all $R$ or $\left( -p\right) $
factors within it are multiplied in the opposite order of the corresponding
operators in $\hat{O}_{R}\left( \hat{x}_{R},\hat{p}_{R}\right) $. The
expressions for $O_{L\star }$ or $O_{\star R}$ can be reduced to a classical
function of ($R^{i},p_{i})$ after using repeatedly the elementary products
given in (\ref{elementaryProducts}) to rewrite $O_{L\star }$ or $O_{\star R}$
as classical expressions $O_{L,R}\left( R,p\right) .$

In the full MSFT only purely $L$ or purely $R$ quantum operators occur, as
above, because of the locality in the $\sigma $ parameter (see footnote (\ref%
{lrproducts})). More generally, in mini-MSFT one may be interested in
writing the QM operator for any Hamiltonian $\hat{H}(\hat{x}_{L},\hat{p}_{L},%
\hat{x}_{R},\hat{p}_{R})$ in the language of star products in iQM. This is
given by representing every elementary $L/R$ operator as left/right star
products according to (\ref{QMrepresentation}),. Hence we get the iQM
representation of any QM Hamiltonian as follows%
\begin{equation}
\hat{H}(\hat{x}_{L},\hat{p}_{L},\hat{x}_{R},\hat{p}_{R})A=H\left( \left(
R\star \right) ,\left( p\star \right) ,\left( \star R\right) ,\left( -\star
p\right) \right) A\left( R,p\right) .  \label{generalH}
\end{equation}%
where the orders of the factors relative to the $\star $s must be preserved.

We give two examples. In the first example we have two particles ($L$ and $%
R) $ interacting with a harmonic oscillator type central force. We can
convert the operator $\hat{H}_{1}$ for this problem to its iQM version by
using the map (\ref{generalH}) that involves only products of string fields 
\begin{align*}
\hat{H}_{1}A& =\left[ \frac{1}{2}\left( \hat{p}_{L}^{2}+\hat{p}%
_{R}^{2}\right) +\frac{\omega ^{2}}{2}\left( \hat{x}_{L}-x_{R}\right) ^{2}%
\right] A\left( R,p\right) \\
& =\frac{1}{2}\left( \vec{p}^{2}+\omega ^{2}\vec{R}^{2}\right) \star A+\frac{%
1}{2}A\star \left( \vec{p}^{2}+\omega ^{2}\vec{R}^{2}\right) -\omega ^{2}%
\vec{R}\star A\star \vec{R} \\
& =\left[ -\frac{1}{4}\hbar ^{2}\partial _{R}^{2}+p^{2}-\frac{\omega ^{2}}{2}%
\hbar ^{2}\partial _{p}^{2}\right] A\left( R,p\right)
\end{align*}%
In the second line only string field products using the string-joining $%
\star $ appear. The last line follows by evaluating the star products by
using (\ref{A12easy},\ref{starLR}). The result in the last line clearly
matches the familiar differential operator representation of the Hamiltonian
as it would be expressed from QM in the $\left( R,p\right) $ basis.

In the second example we illustrate the Hamiltonian $\hat{H}_{2}$ derived
from string theory in 2-dimensions with quarks (0-branes) attached at the
ends \cite{2Dstring}, where the positions of the quarks $x_{L,R}^{\mu }$ (in
the lightcone basis) are actually the end points of the string, 
\begin{eqnarray}
\hat{H}_{2}A &=&\left[ \frac{m_{L}^{2}}{2\hat{p}_{L}^{+}}+\frac{m_{R}^{2}}{2%
\hat{p}_{R}^{+}}+\gamma \left\vert \hat{x}_{L}^{-}-\hat{x}%
_{R}^{-}\right\vert \right] A\left( R,p\right)  \notag \\
&=&\frac{m_{L}^{2}}{2p}\star A+A\star \frac{m_{R}^{2}}{2p}+\gamma \left\vert
\left( R\star \right) -\left( \star R\right) \right\vert A\left( R,p\right) 
\notag \\
&=&\frac{m_{L}^{2}}{2p}\star A\left( R,p\right) +A\left( R,p\right) \star 
\frac{m_{R}^{2}}{2p}+\gamma \hbar \int^{\prime }\frac{dk}{\pi k^{2}}A\left(
R,p+k\right)  \label{2dstring}
\end{eqnarray}%
In the second line the map (\ref{generalH}) is used to connect the $\star $
version to the QM\ operator version. In the third line the prime on $%
\int^{\prime }$ means the principal value integral which arises from
computing the star products in the second line. The last line reproduces
exactly the spectrum of large-N QCD in two dimensions ('t Hooft's integral
equation for a meson \cite{thooft}), as expected from \cite{2Dstring}, but
we will skip the details here\footnote{%
In \cite{thooft}\cite{2Dstring} the wavefunction is in momentum space $%
\left( p_{L},p_{R}\right) $, whereas in (\ref{2dstring}) it is in the mixed
phase space $A\left( R,p\right) .$ After a Fourrier transform ($R\rightarrow
P$) and appropriate change of variables from $\left( P,p\right) $ to $\left(
p_{L},p_{R}\right) $ we find the same integral equation for mesons.}.

For any choice of Hamiltonian we can define the quadratic term of the field
theory for the mini-MSFT, and furthermore we can include \textquotedblleft
string\textquotedblright -\textquotedblleft string\textquotedblright\
interactions by imitating MSFT as follows%
\begin{equation}
S=\int d^{n}Rd^{n}p\left[ \frac{1}{2}A\left( \hat{H}A\right) +\frac{g}{3}%
A\star A\star A+\cdots \right] .  \label{Smini}
\end{equation}%
Here the dots $+\cdots $ imply that many mini-MSFT models may be constructed
that include higher powers of field interactions beyond the cubic term. The
Feynman-like diagrams for this field theory reproduce the joining/splitting
of worldsheets as in the old string-like \textquotedblleft duality
diagrams\textquotedblright . We think that with only the cubic interaction
in (\ref{Smini}), and the 2D string Hamiltonian of Eq.(\ref{2dstring}), it
seems that the mini-MSFT approach would parallel the 2D string Feynman
diagram computations in (\cite{2Dstring}) that gave correctly the meson
interactions by using only strings and branes (quarks at the end) with
amplitudes in agreement with planar graphs in 2D large N QCD. Perhaps this
successful and exact string-QCD correspondence could now be generalized to
four dimensions through mini-MSFT in (\ref{Smini}) by including the
transverse components of $R^{\mu },p_{\mu }$ beyond the lighcone components.

This completes the construction of the mini-MSFT field theory model. Time
will show if this is a useful approach to discuss some physical systems,
such as QCD strings. It is possible to generalize the system further by
allowing $A$ to carry labels that correspond to spin and other quantum
numbers and correspondingly choose an appropriate $\hat{H}$. In this paper
the mini-MSFT concept was used mainly as a simplification of the full MSFT
to discuss the link between the string-joinning star product and the
quantization rules of QM. As shown in \cite{MSFT-BR1} all facets of our
discussions here are also true in the full MSFT as well as subsectors,
derivable from it.

\section{Outlook}

We have shown that in the half-phase-space of iQM we can reproduce all
aspects of ordinary QM by relying only on the rules of the $\star $ product
whose physical meaning is the interactions created by joining/splitting
strings.

To make the final point for the central thesis in this paper regarding the
source for the rules of quantum mechanics in all physics, morally one needs
to first imagine that string theory (or the M-theory generalization) may
indeed be the correct description for all physical phenomena. Then based on
the full MSFT \cite{MSFT-BR1} (and its potential generalization to
M-theory), one may claim that the source of quantum mechanical commutation
rules in all physics could be traced back to the physical phenomenon of
string joining/splitting interactions as expressed in the half-phase-space
in the MSFT language. If this view holds up beyond the apparent limited
reach of MSFT into all aspects of M-theory, including second quantization,
then the concept we discussed here for string interactions being the source
for quantum mechanics would boosts the credibility of string theory as a
fundamental theory.

Let us re-assess the main ingredients that yield these results. First, there
is the iQM generated by the string-joining $\star $ product of Eq.(\ref%
{moyal}) which comes from the corresponding Eq.(\ref{msftstar}) in full
string field theory. Second there is the connection of the quantum operators
in QM to the string joining star product as given in Eq.(\ref%
{QMrepresentation}), which also comes from the full string field theory in
Eq.(\ref{iQMX2},\ref{iQMP2}). The second ingredient may be regarded as a
particular \textit{representation of the quantum operators.} Of course,
being a representation, it is bound to satisfy the correct quantum rules. In
fact, the string version in Eq.(\ref{iQMX2},\ref{iQMP2}) was first arrived
at in \cite{MSFT-BR1} from the study of the quantized string, namely from
the knowledge acquired in QM. What is new is that unlike other
representations, this representation is based on a physical process of
string joining/splitting that takes place at the Planck scale. In other
words, while being a representation it is also connected to physical
processes in a way that other representations of quantum operators are not.
This provides the seed of an explanation that quantum mechanics exists
because of certain phenomena, while other representations do not have this
capacity. In this representation, if the physical processes of string
joining/splitting do not occur, there is no quantum mechanics, because the
non-commutativity parameter in string joining splitting is none other than
the Planck constant $\hbar $. Therefore, we reverse the logical path that
brought us from quantum mechanics, through string theory, to SFT in
particular MSFT. We consider the premise that string theory or M-theory is
the primary starting point for the description of all phenamena in nature.
This requires that there are no point like objects, that all objects are
fundamentally string-like and that they must interact only via the process
of string joining/splitting. The language of MSFT makes it clear that in
that case an induced quantum mechanics arises, and that the $\hbar $ of
quantum mechanics comes from the non-commutativity of string
joining/splitting. In this view we may say that the standard QM operators
and the corresponding commutation rules are introduced for convenience
through Eqs.(\ref{iQMX2},\ref{iQMP2},\ref{QMrepresentation}) in order to
make a connection to familiar language, but not as fundamental, and also not
because they are needed in order to compute - MSFT is already equipped with
the tools of computation.

Independent of the central thesis in this paper, at a more modest level, we
have introduced a new representation space for the quantum mechanical
operators through the map in Eq.(\ref{QMrepresentation}) which may find many
applications. The mini-MSFT model may be useful in its own right to discuss
some perturbative and non-perturbative physics in certain circumstances.

\begin{acknowledgments}
This research was partially supported by the U.S. Department of Energy. IB
thanks CERN for hospitality while part of this research was performed.
\end{acknowledgments}

\end{document}